\documentclass{appolbnh}
\usepackage{graphicx}

\begin{document}

\title{Revisiting $KN$ elastic and charge exchange reactions in search of the pentaquark $\Theta^+$ baryon}

\author{Byung-Geel Yu\thanks{bgyu@kau.ac.kr}, Kook-Jin Kong\thanks{kong@kau.ac.kr}
\address{Research Institute of Basic Science, Korea Aerospace University, Goyang, 10540, Korea}
\\[3mm]
{Tae Keun Choi\thanks{tkchoi@yonsei.ac.kr}
\address{Department of physics and Engineering Physics, Yonsei University, Wonju, 26493, Korea}}
}

\maketitle

\begin{abstract}

In the search for the pentaquark $\Theta^+$ baryon, the world data on $K^+n\to K^+n$ and $K^+n\to K^0\,p$ reactions on the deuteron target and $K^0\,p\to K^+n$ and $K^0_L\,p\to K^0_S\,p$ reactions on the hydrogen target are revisited to study the isoscalar component of the scattering amplitudes.
The determination of the $s$- and $p$-wave phase shifts for the two former channels on the deuteron target is presented in the low momentum region with the uncertainty due to the deuteron form factors at forward angles discussed.
For the reactions on hydrogen targets, a similarity expected from time reversal between the charge exchange processes $K^+n\to K^0\,p$ and $K^0\,p\to K^+n$ is exploited, while we present the analysis of the $K_L^0\,p\to K_S^0\,p$ scattering based on the $s$-wave phase shift for low momenta combined with the $t$-channel $\rho^0(775)+\omega(782)+\phi(1020)$ vector meson exchange at high momenta.
With the mass 1535 MeV and decay width 5 MeV the possibility of the $\Theta^+$ in the Breit-Wigner (BW) form is tested in the differential cross section measured by C. J. S. Damerell {\it et\ al.} at ${\rm P_{Lab}}=434$ MeV/$c$ and polarization.
Our analysis suggests measuring the polarization observable rather than the differential cross section, which has sufficient sensitivity to distinguish the $\Theta^+$ baryon among all spin and parity states $J^P = 1/2^\pm$ and $3/2^\pm$.

\end{abstract}

\section{Introduction}

Since M. Gell-Mann's original idea of the exotic configuration of quark combinations \cite{gellmann}, it has been anticipated that the multiquark structures of hadrons, $\{q\bar{q}q\bar{q}\}$ and $\{qqqq\bar{q}\}$ for instance, can contribute to deepening our understanding of the strong force.
Exploring these structures may lead to significant revisions in our understanding of how quarks combine and interact, potentially revealing new aspects of the fundamental forces that govern our universe.

D. Diakonov, V. Petrov and M. V. Polyakov made a theoretical prediction for the existence of the pentaquark $\Theta^+$ baryon of isosinglet strangeness $+1$ particle with mass $1530$ MeV, decay width $\Gamma_{\Theta^+}\leq15$ MeV, and spin $1/2$, which arises from the chiral soliton picture \cite{diakonov}.
Since the publication of this memorial work and the subsequent study by M. Prasza${\l}$owicz \cite{prasz}, there has been a renewed interest in studying the internal structure of hadrons. Specific studies of the exotic $\Theta^+$ have been carried out both in theory and experiment, in particular in the photoproduction reaction $\gamma N\to KKN$ and $KN$ scattering from which the $\Theta^+$ can be observed.

The empirical verification of the $\Theta^+$ was attempted by T. Nakano ${et\ al.}$ in the photoproduction $\gamma N\to \Theta^+K\to NKK$ experiment at the LEPS Collaboration \cite{nakano}, claiming that the exotic baryon is observed in the reaction on the $^{12}C$ target at $1.54\pm0.01$ GeV with $\Gamma_{\Theta^+}\leq25$ MeV, though more rigorous verification awaits since then.
%
%
On the theoretical side, questions concerning the determination of the spin parity of the $\Theta^+$ baryon in the photoproduction process have been investigated in hadron models \cite{bgyu1,bgyu2}.
Meanwhile, experiments in search of an exotic baryon of $s=+1$ were carried out in the 1970s in $KN$ scattering to look for the resonance called $Z^*$ with an assumed mass of 1750 MeV and a decay width in the range 300 $\sim$ 600 MeV, respectively \cite{bgrt,hirata,carroll}. Therefore, the formation of $Z^*$ in experiments was sought in measurements at kaon momentum ${\rm P_{Lab}}\approx 900$ MeV/$c$.

According to Ref. \cite{diakonov}, the mass and width of the resonance have now been predicted to be smaller than expected in previous experiments. In particular, the $K^+n\to K^+n$ and $K^+n\to K^0p$ scattering experiments conducted by Damerell {\it et\ al.} \cite{damerell}, G. Giacomelli {\it et\ al.} \cite{giacomelli72,giacomelli73,giacomelli74}, and R. G. Glasser {\it et\ al.} \cite{glasser} approached the $\Theta^+$ resonance point ${\rm P_{Lab}}\approx430$ MeV/$c$, anticipating the possibility of finding the pentaquark configuration $\{uudd\bar{s}\}$ with the initial state kaon beam of $s=+1$.
Recently, to study the feasibility of the $\Theta^+$ in the $K^+d\to K^0\, pp$ reaction, the invariant mass spectrum $K^0p$ corresponding to the $\Theta^+$ mass is analyzed in the reaction cross section \cite{sekihara}. The results provide a theoretical background on which the empirical search for the $\Theta^+$ at J-PARC is proposed \cite{jkahn}.
On the other hand, the $K^0\,p$ scattering on the hydrogen target, which can compensate for the uncertainty in the $K^+n$ scattering due to the deuteron form factor, should be of interest, although experiments have difficulty in dealing with $K^0_L$ and $K^0_S$ components which weakly decay to multipions during the reaction process.
%
Analysis of the $K^0\,p\to K^+n$ and $K^+n\to K^0\,p$ reactions related by time reversal will provide insight into the differences between free and bound neutrons, which could help to understand the complexity of the reactions on the deuteron target.
Resuming the experiments on the $K^0p\to K^+n$ reaction measured typically by J. C. M. Armitage {\it et\ al.} \cite{armitage} in the 1970s, more recently, the Hall D Collaboration plans to measure this experiment anew with an improved apparatus in the KLF experiment at JLab \cite{amaryan24}. Also the experiment on the $K^0_L \,p\to K^0_S\, p$ scattering conducted by G. W. Brandenburg {\it et\ al.} \cite{brandenburg} could offer a chance to extract the information about the $\Theta^+$ based on the cleaner background. Nevertheless, due to the limited momentum range down to 1 GeV/$c$, there are no data near ${\rm P_{Lab}}\approx 410$ MeV/$c$ at which the possibility of the $\Theta^+$ with its mass 1530 MeV could be investigated.

At this stage, however, it is worth remarking that in contrast to these studies and their achievements, experiments denying the possible existence of the $\Theta^+$ baryon have been reported almost equally, as a highly statistical experiment at JLab failed to confirm previous claims of success, concluding that the pentaquark either does not exist at all or has a very small cross section and is currently unobservable \cite{amaryan22}.

Keeping in mind the current progress on the issue of the pentaquark $\Theta^+$, we investigate the possibility of the $\Theta^+$ based on $KN$ scattering with the motivation that previous experiments conducted for this purpose need to revisit this question around the kaon momentum of 434 MeV/$c$ \footnote{There were experimental data on $K^+n\to K^0\,p$ and $K^+n\to K^+n$ reactions at the momentum measured by Damerell {\it et\ al.}.} in accordance with the mass 1540 MeV and the smaller decay width.
In this work, we extend our previous study of the possibility of the $\Theta^+$ in the $K^+n$ scattering \cite{ky-kn} to include the $K^0\,p$ channels with the decay width $\Gamma_{\Theta^+}=5$ MeV chosen in our theoretical calculation to be consistent with experiments.
The first half of the paper contains a revision of our previous work, while the remaining part further investigates the possibility of the exotic $\Theta^+$ in the $K^0\,p\to K^+n$ and $K^0_L\,p\to K^0_S\,p$ channels. Since the experimental data are available for the latter process up to ${\rm P_{Lab}}\approx10$ GeV/$c$, we present a description of the Regge pole exchange for the scattering cross section at high momenta to provide a theoretical framework for future studies.

The paper is organized as follows: In Section 2, we address the uncertainty arising from the deuteron form factor. We determine the isoscalar amplitude of $K^+n\to K^+n$ and $K^+n\to K^0p$ reactions, by establishing the isovector amplitude from the $K^+p\to K^+p$ scattering.
Section 3 is devoted to the numerical analysis of the possibility of the exotic $\Theta^+$ in these $K^+n$ reactions. Total and differential cross sections including polarization observables are investigated based on the phase shift of $s$- and $p$-waves with the $\Theta^+$ given in the BW form.
In Sec. 4 we explore the reaction $K^0\,p\to K^+n$ based on the validity of time reversal for the inverse channel $K^+n\to K^0\,p$. We also discuss the $K^0_L\,p\to K^0_S\,p$ scattering within the framework of the $s$-wave phase shift, incorporating the $t$-channel vector meson exchange.
A summary and conclusion for the exotic $\Theta^+$ search in $KN$ scattering is given in Sec. 5.

\section{$KN$ scattering on deuteron target}

\subsection{Partial wave analysis for $K^+N$ reactions}

We introduce the following four scattering channels in the $KN$ reaction to probe the existence of the exotic $\Theta^+$ baryon with kaon beams;
\begin{eqnarray}
&&K^+n\to K^+n\,,\label{k+n}\\
&&K^+n\to K^0p\,,\label{k+ncex}\\
&&K^0p\to K^+n \,,\label{k0pcex}\\
&&K^0p\to K^0p \,,\label{k0p}
\end{eqnarray}
where the initial $K^0$ is understood to be in a $K^0_L$ state, long-lived over the $K^0_S$ state.

We are interested in determining the above $KN$ scattering amplitudes, starting with a simple fit of the parameters for the $s$-wave phase shift in the $K^+ p$ elastic reaction in the low momentum region.
We note that a similarity is expected due to time reversal between Eqs. (\ref{k+ncex}) and (\ref{k0pcex}) and the isospin invariance of both the reactions \footnote{
It is also called charge symmetry in the sense that it is invariant when the $u$ and $d$ quarks are swapped in the other process with electromagnetic and mass difference neglected. This should not be confused with the symmetry with respect to charge conjugation.} between Eqs. (\ref{k+n}) and (\ref{k0p}).
Thus, the pairs of corresponding amplitudes share the same expression for the isospin combination as in Eqs. (\ref{k+ncex-amp}) and (\ref{k0pcex-amp}), and Eqs. (\ref{k+n-amp}) and (\ref{k0p-amp}) below.
In addition to the advantages and disadvantages of reactions on deuteron and hydrogen targets mentioned above, it is likely that such a similarity, if observed, will provide further information on the reaction mechanism involved.

In this section, we will determine the isovector and isoscalar components of the $K^+n$ reactions on the deuteron and apply the results to the amplitudes of the $K^0\,p$ reactions on a hydrogen target with our interest in evaluating the validity of such a similarity between the two different targets.

\subsection{The scenario for the determination of the isoscalar amplitude }

Since the isospin of the kaon and nucleon is 1/2, the scattering amplitudes can be expressed as the sum of the isoscalar($I=0$) and isovector($I=1$) amplitudes,
\begin{eqnarray}
&&{\cal M}(K^+n\to K^+n)={1\over2}\left({\cal M}^{(1)}+{\cal M}^{(0)}\right)
\label{k+n-amp},\\
&&{\cal M}(K^+n\to K^0p)={1\over2}\left({\cal M}^{(1)}-{\cal
M}^{(0)}\right)\label{k+ncex-amp}\,,\\
&&{\cal M}(K^0p\to K^+n)
={1\over2}\left({\cal M}^{(1)}-{\cal M}^{(0)}\right)\label{k0pcex-amp}\,,\\
&&{\cal M}(K^0p\to K^0p)
={1\over2}\left({\cal M}^{(1)}+{\cal M}^{(0)}\right)\label{k0p-amp}\,,
\end{eqnarray}
where ${\cal M}^{(0)}$ and ${\cal M}^{(1)}$ are the isoscalar and isovector components of the reaction amplitude common to all these channels.

Then the following is our scenario for determining the amplitudes of $K^+n$ and $K^0p$ scattering above; in order to fix the isovector amplitude first, we utilize the scattering amplitude for the elastic channel $K^+p\to K^+p$, which consists of
\begin{eqnarray}
{\cal M}(K^+p\to K^+p)={\cal M}^{(1)}+{\cal M}_C\,\label{k+p-amp}
\end{eqnarray}
with the Coulomb amplitude ${\cal M}_C$ due to the repulsive interaction between $K^+$ and proton \cite{aoki,hashimoto}.
For this purpose, we use the phase shift analysis of the experimental data in Eq. (\ref{k+p-amp}) based on the partial wave expansion since any hadron model such as in Ref. \cite{yk-kn} for the $t$ channel meson exchange is invalid for the reaction in the low momentum region $p_{K}\leq3$ GeV/$c$.
Given the isovector amplitude ${\cal M}^{(1)}$ thus determined from the step in Eq. (\ref{k+p-amp}), the isoscalar amplitude ${\cal M}^{(0)}$ remaining in Eqs. (\ref{k+n-amp}) and (\ref{k+ncex-amp}) must be constructed to agree with the empirical data on total and differential cross sections at low momenta.

Before proceeding, it is mandatory to study the effect of the bound state of the nucleon in the deuteron when dealing with the kaon interaction with the neutron target inside.

\subsection{Deuteron form factors}


\begin{figure}[]
\centerline{
\includegraphics[width=8cm]{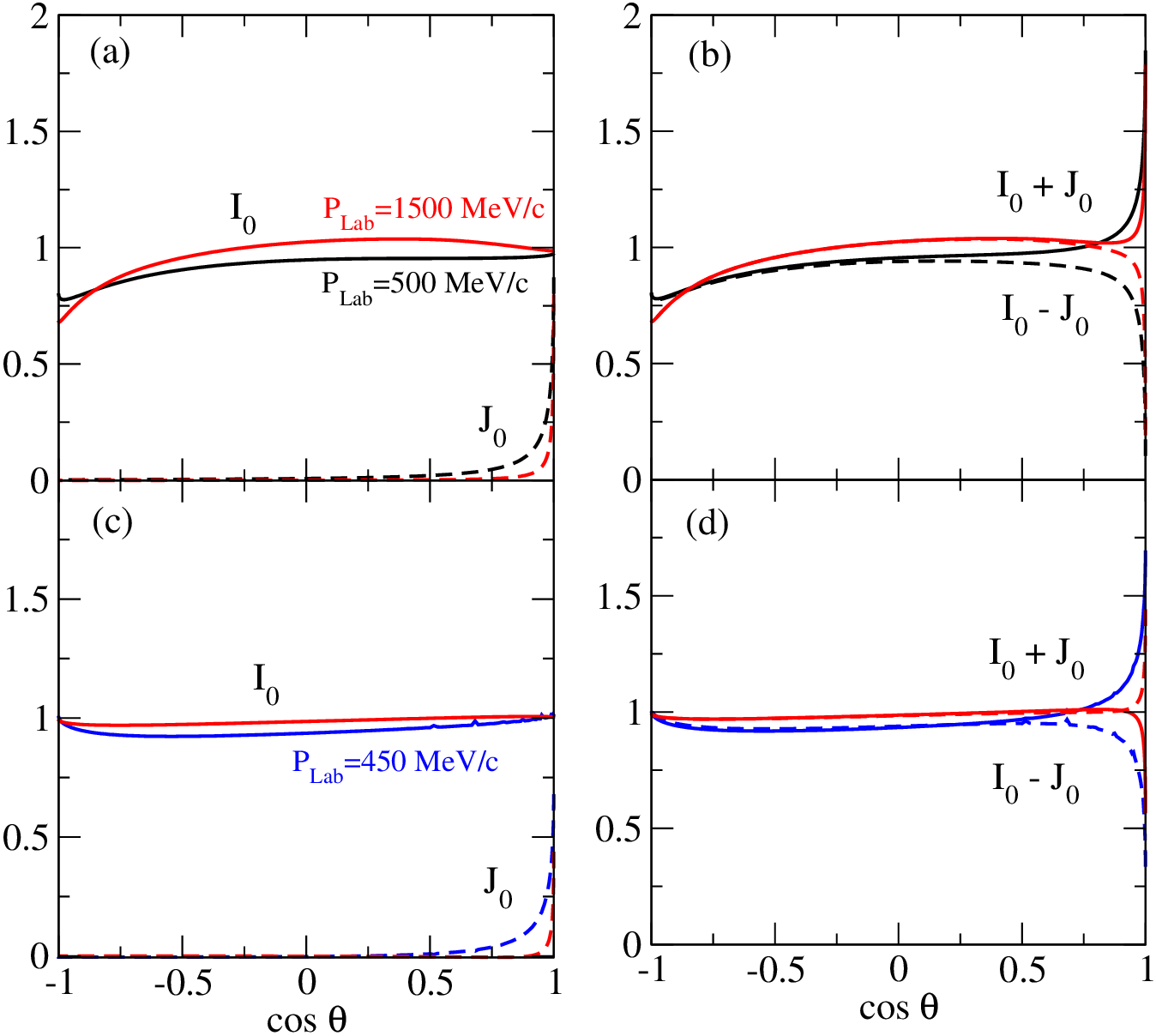}
}
\caption{Deuteron form factors $I_0(\theta)$ and $J_0(\theta)$
vs c. m. angle $\theta$ at the kaon lab momentum. The Hulthen wave function is used in (a) and (b) for the
delta function integration method with the cutoff $\Lambda=200$
MeV applied. In (c) and (d) the Moravcsik-Gartenhaus wave function
\cite{giacomelli73} is used with the same cutoff $\Lambda$.}
\label{fig1}
\end{figure}

The experimental data on kaon scattering off a neutron target are obtained from the scattering off a deuteron target $K^+d \to K^+n(p)$, or $K^+d \to K^0p(p)$ with the spectator proton \cite{damerell,giacomelli73,sekihara}.
Thus, the reaction cross sections should take into account the deuteron inelastic form factors $I_0$ and $J_0$ \cite{glasser,hashimoto},
\begin{eqnarray}
&&{d\sigma\over d\Omega}\left[K^+d\to K^+n(p)\right]=\left(\left|f_n\right|^2+\left|g_n\right|^2\right)I_0,\label{d-k+n}\\
&&{d\sigma\over d\Omega}\left[K^+d\to K^0p(p)\right]=\left(\left|f_{\rm cx}\right|^2+{2\over3}\left|g_{\rm cx}\right|^2\right)\left(I_0-J_0\right)
+{1\over3}\left|g_{\rm cx}\right|^2\left(I_0+J_0\right),\label{d-k0p}\,
\nonumber\\
\end{eqnarray}
where $f_n\,(g_n)$ is the spin non-flip (spin flip) elastic scattering amplitude, introduced as ${\cal M}^{(I)}=f^{(I)}+i\vec\sigma\cdot\hat{n}\,g^{(I)}$, and similarly $f_{\rm cx}\,(g_{\rm cx})$ for the charge exchange process from the neutron target. The inelastic form factors $I_0$ and $J_0$ represent the scattering from a deuteron, containing information on the deuteron internal structure and the Pauli exclusion principle which invalidates the measurements at small scattering angles by suppressing the spin non-flip process \cite{edelstein};
\begin{eqnarray}
&&I_0(\theta)=K\int{k^2dk\over \omega}{d^3p\over E_p}{d^3q\over E_q}\delta^4(k_0+P-k-p-q)
{1\over2}\left[u^2(p)+u^2(q)\right],\ \  \\
&&J_0(\theta)=K\int{k^2dk\over \omega}{d^3p\over E_p}{d^3q\over E_q}\delta^4(k_0+P-k-p-q)
u(p)u(q)
\end{eqnarray}
with the deuteron wave function $u(p)$ in momentum space and details of the kinematic constant $K$, kaon and nucleon energies $\omega$, $E_p$ and $E_q$ given in Ref. \cite{hashimoto}.

Figure \ref{fig1} illustrates the dependence of form factors $I_0$ and $J_0$ on the scattering angle $\theta$ in the c. m. frame at the incident kaon momentum ${\rm P_{Lab}}$ in the laboratory frame.
From the form factors predicted at ${\rm P_{Lab}}=450$ and 1500 MeV/$c$ with the chosen cutoff mass $\Lambda=200$ MeV, it is sufficient to consider that $I_0\approx1$ and $J_0\approx0$, except for the non-vanishing contribution of $J_0$ and $I_0\pm J_0$ in the forward and backward directions.
Thus, avoiding the measurement at forward and backward angles, these form factors lead to the $K^+n$ scattering amplitudes in Eqs. (\ref{d-k+n}) and (\ref{d-k0p}) of the form $|f|^2+|g|^2$ like a free neutron target.
It should be noted, however, that the effect of the form factors is not negligible at forward angles $\cos\theta\geq 0.7$, which prevents a clear observation of such a tiny resonance peak as the $\Theta^+$ in the forward direction.
This would be the reason why some of the experiments on the deuteron target report a negative result in the search for $\Theta^+$ at forward angles.

With these in mind we now determine the isovector amplitude ${\cal M}^{(1)}$ in Eq. (\ref{k+p-amp}) from the experimental data.

\subsection{Isovector amplitude}

In the elastic $K^+p\to K^+p$ scattering below ${\rm
P_{Lab}}\approx 800$ MeV/$c$, the total cross section is almost
constant in this region. Apart from the Coulomb repulsion, it is due to the repulsive
hadronic interaction between $K^+$ and the proton, which gives an
indication of the $s$-wave phase shift.
In the differential cross section, the angular dependence is
isotropic except for the sharp peaks at very forward angles
due to the Coulomb repulsion.

To implement such isotropy in the differential cross section, a
partial $s$-wave with the phase shift is considered. Denoted by
the symbol $S_{11}$, the isovector amplitude for the $s$-wave is
written as \footnote{The kinematic factor $C$ is called for
the correct dimension of the partial wave amplitude following the
CGLN notation to the meson-baryon scattering,
\begin{eqnarray}
C={8\pi W\over\sqrt{4MM'}}\sqrt{{1\over k\,q}}\ ,\nonumber
\end{eqnarray}
where $k$ and $q$ are the initial and final
kaon momenta in the c. m. frame and $M(M')$ is the initial(final)
nucleon mass.}
\begin{eqnarray}\label{phase-s11}
S_{11}={1\over2ik}
\left(\eta^1_{0+} e^{2i\delta^1_{0+}}-1\right), 
\end{eqnarray}
where the inelasticity $\eta^1_{0+}=1$ is assumed without loss of generality.

The phase shift of $S_{11}$ is obtained as a linear function of
the incident kaon momentum $k$ in the c. m. frame, i.e.,
\begin{eqnarray}
&&\delta^1_{0+}(k)=a_0+b_0 k \label{goldhaber}
\end{eqnarray}
with the coefficients $a_0=3$ and $b_0=-107$ GeV$^{-1}$ fixed to
the differential cross section data. The
phase shift is negative (see the set I and set II in Fig. \ref{fig3} right panels below), consistent
with the repulsive interaction between $K^+$ and the
proton. Our fit in Eq. (\ref{goldhaber}) is almost the same as
that of Goldhaber \cite{goldhaber}, and hence the full amplitude
is given by
\begin{eqnarray}\label{s11}
{\cal M}(K^+p\to K^+p) = S_{11}+{\cal M}_C
\end{eqnarray}
with the Coulomb interaction term ${\cal M}_C$ discussed in detail in Ref. \cite{aoki}.
As expected, the Coulomb interaction is restricted to a very short range of angles.

With the result of Eq. (\ref{s11}), we can now construct partial waves for
the isoscalar amplitude ${\cal M}^{(0)}$ in Eqs. (\ref{k+n-amp}) and (\ref{k+ncex-amp}).

\subsection{Isoscalar amplitude}

Now that we are dealing with the low momentum reaction below 800 MeV/$c$,
it is good to consider the $s$- and $p$-waves for the isoscalar amplitude.
Similar to the isovector case above, the isoscalar $s$-wave amplitude is written as
\begin{eqnarray}\label{phase-s01}
S_{01}={1\over2ik}
\left(\eta^0_{0+} e^{2i\delta^0_{0+}}-1\right),
\end{eqnarray}
and the partial $p$-waves are further constructed as,
\begin{eqnarray}
&&P_{01}=f^0_{1-}\cos\theta-i\vec{\sigma}\cdot\hat{n}f^0_{1-}\sin\theta,\\
&&P_{03}=2f^0_{1+}\cos\theta+i\vec{\sigma}\cdot\hat{n}f^0_{1+}\sin\theta,
\end{eqnarray}
where
\begin{eqnarray}\label{p1}
f^0_{1\pm}={1\over2ik}\left(\eta^0_{1\pm} e^{2i\delta^0_{1\pm}}-1\right)
\end{eqnarray}
with $\eta^0_{0+}=1$ and $\eta^0_{1\pm}=1$ as before.

\begin{figure}[b]
\centerline{
\includegraphics[width=6cm]{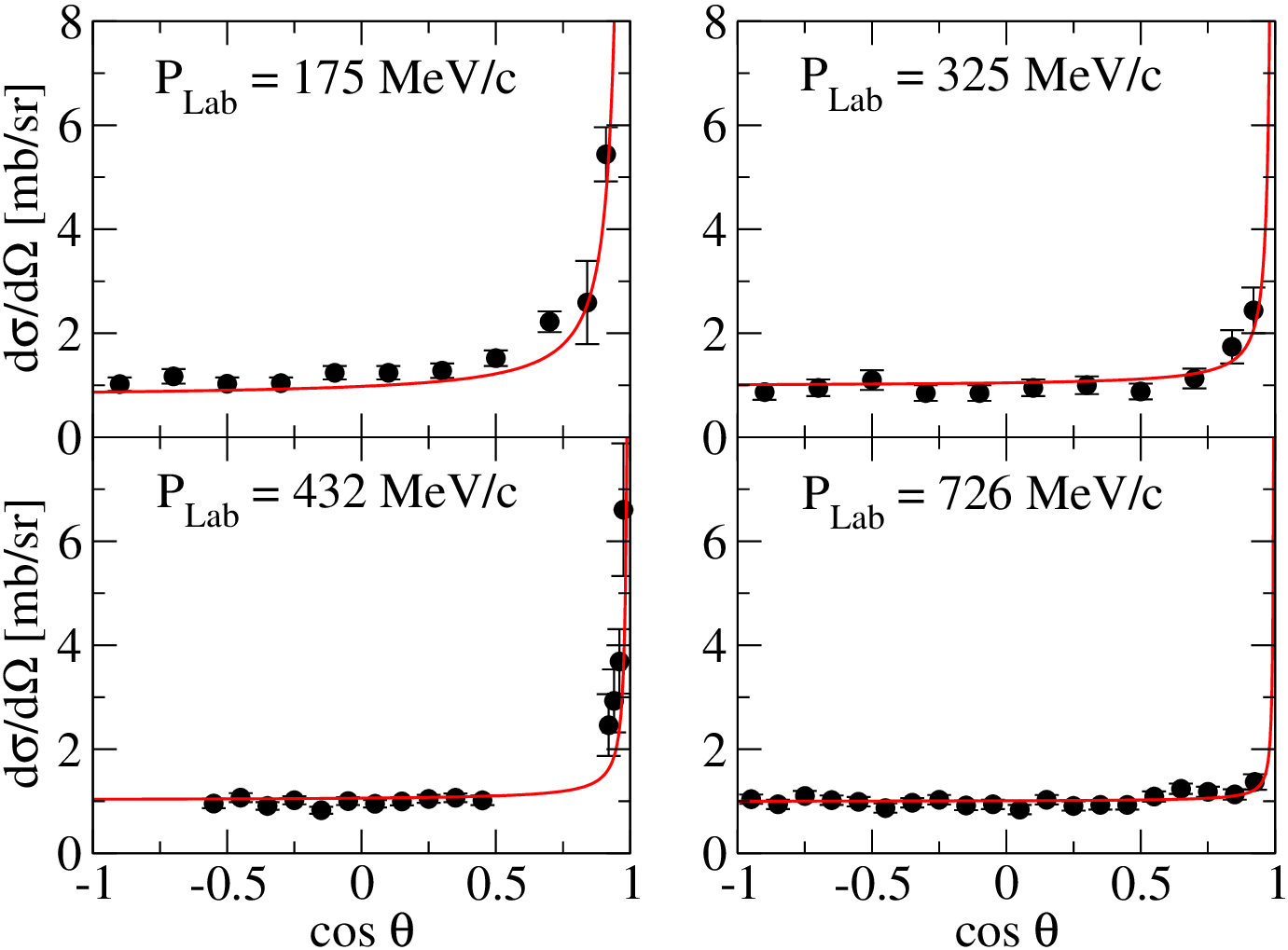}\\
\hspace{0.2cm}
\includegraphics[width=6cm]{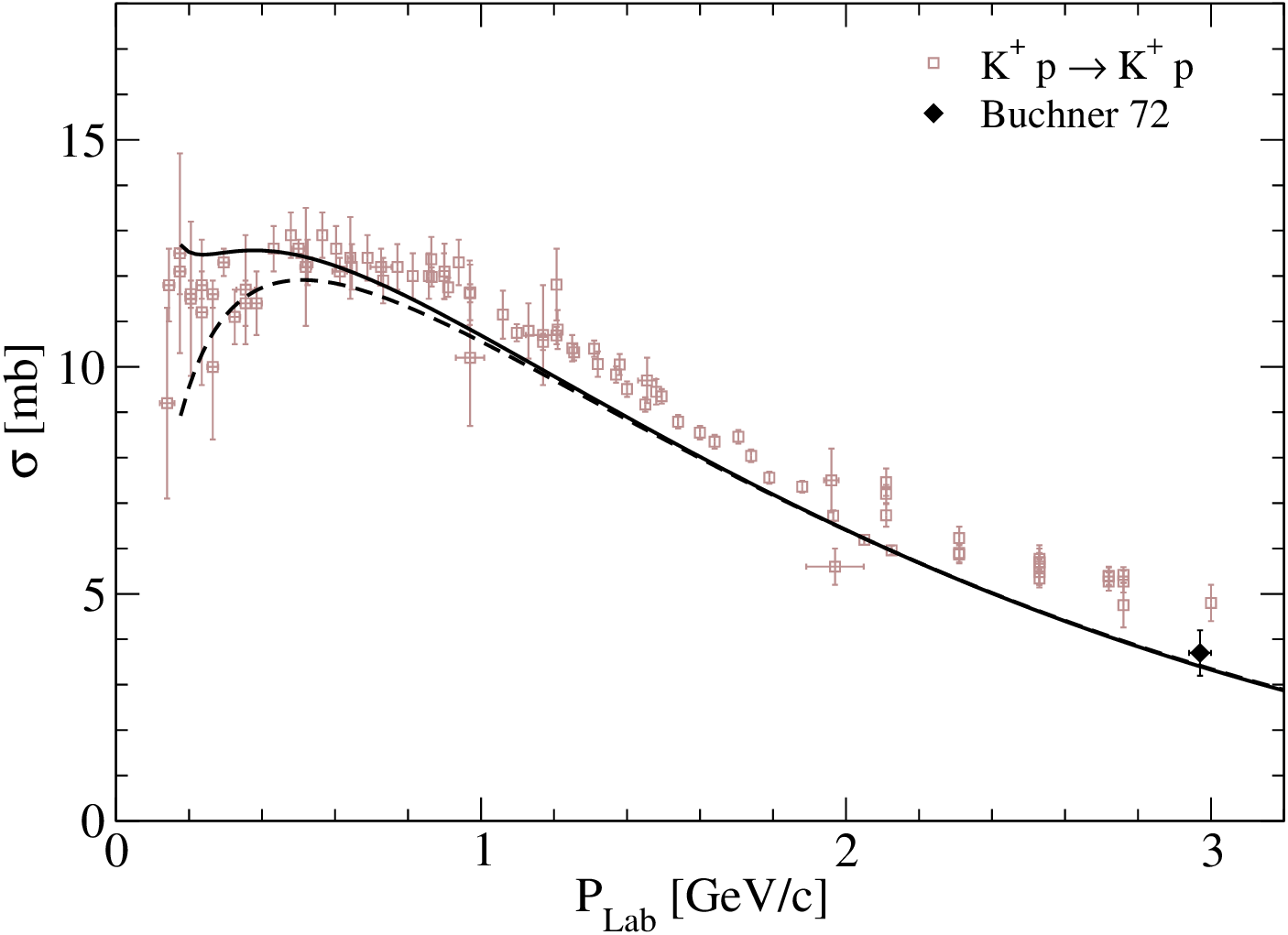}
}
\caption{Differential and total cross sections
for elastic $K^+p\to K^+p$ scattering from Eq. (\ref{s11}). The Coulomb repulsion is
responsible for the forward peaks in the differential cross sections.
Data for the differential cross section are taken from Ref. \cite{cameron}.
The solid curve for the total cross section includes the Coulomb effect and the dashed one without it.
Data are from Particle Data Group with the point at ${\rm P_{Lab}}=2.97\pm0.03$ GeV taken from Ref. \cite{buchner}.
} \label{fig2}
\end{figure}

Thus, from the isospin relations in Eqs. (\ref{k+n-amp}),
(\ref{k+ncex-amp}), (\ref{k+p-amp}) and (\ref{s11}) above,
the scattering amplitudes for $K^+n \to K^+n$ and
$K^+n \to K^0p$ reactions are written in terms of these $s$- and $p$-waves,
\begin{eqnarray}
&&{\cal M}(K^+n \to K^+n)={1\over2} \left(S_{11}+S_{01}+P_{01}+P_{03}\right),
\label{phase-1}\\
&&{\cal M}(K^+n \to K^0p)={1\over2} \left(S_{11}-S_{01}-P_{01}-P_{03}\right).
\label{phase-2}
\end{eqnarray}

For the elastic and charge exchange $K^+n$ reactions mentioned above, there are two sets of differential cross section data measured by Damerell {\it et\ al.} \cite{damerell} and by Giacomelli {\it et\ al.} \cite{giacomelli73}, both of which share a common range of kaon momentum from 640 to 940 MeV/$c$.
Thus, there are two possible approaches in this overlapping region. We will focus on Damerell's data first to determine the isoscalar amplitudes $S_{01}$, $P_{01}$ and $P_{03}$ in Eqs. (\ref{phase-1}) and (\ref{phase-2}).
We refer to this as set I. The other then involves fitting Giacomelli's data, which we call set II \cite{ky-kn}.

\section{Numerical results}

\subsection{Isovector amplitude}


Given the phase shift $\delta^1_{0+}(k)$ for the isovector amplitude $S_{11}$ in Eq. (\ref{goldhaber}), the differential cross sections $d\sigma/d\Omega$ for the elastic $K^+p$ reaction in the range $150 \leq {\rm P_{Lab}} \leq 750$ ${\rm MeV}/c$ are shown in Fig. \ref{fig2}.
The isotropic pattern is clearly visible, except for the Coulomb repulsion which peaks sharply at the very forward angles.
The total cross section with and without Coulomb repulsion is shown by the solid and dashed curves, respectively.
Since it is highly divergent at the angle $\theta\to 0$, we obtain the total cross section by restricting the range of the angel to $-1<\cos\theta<0.85$ in the integration of the differential cross section.
As the $s$-wave with the phase shift linear in $k$ in Eq. (\ref{goldhaber}) reproduces the total cross section to some extent, the $s$-wave dominance assumed in the fit is plausible in the low momentum region, even though the anisotropy of the differential cross section becomes stronger above the ${\rm P_{Lab}}\geq 800$ MeV/$c$.

\subsection{Isoscalar amplitude}

\begin{figure}[]
\centerline{
\includegraphics[width=6cm]{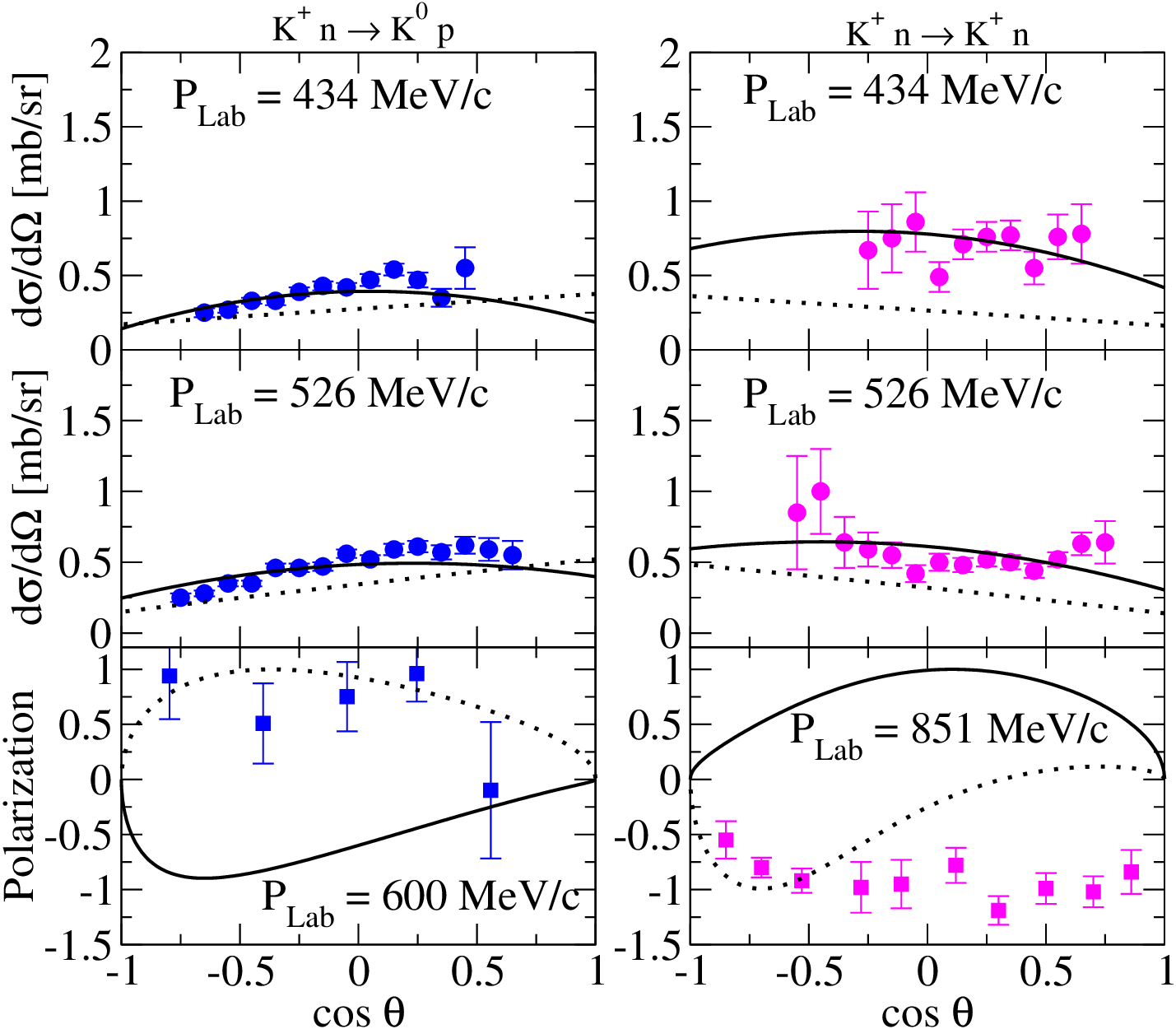}\\
\hspace{0.0cm}
\includegraphics[width=6.5cm]{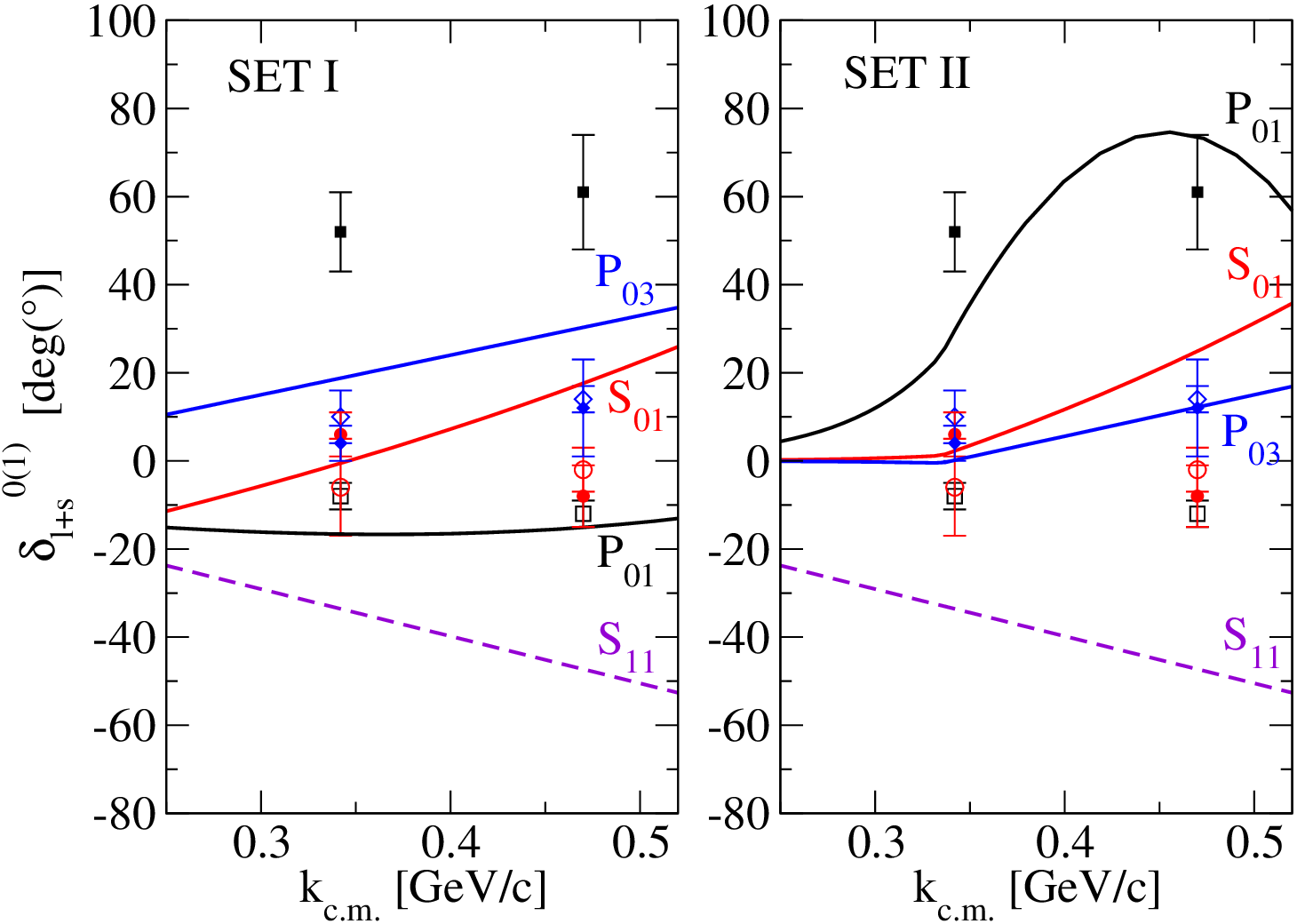}
}
\caption{Differential cross sections and polarizations
for $K^+n\to K^0p$ and $K^+n\to K^+n$ (left) and the phase shift from set I and set II (right).
In the left; the solid curve results from set I and the dotted one from set II, respectively.
Differential cross section data favor the parameters from set I, whereas the polarization follows set II.
Data are taken from Ref. \cite{damerell} for differential cross section and Refs. \cite{ray,Robertson,Watts,Nakajima} for polarization. %
In the right;
the phase shift $\delta_{l+s}^{0(1)}$ with the notation $L_{I\,2J}$ from set I (left) and set II (right) is presented in the kaon momentum range 0.25 $\sim$ 0.55 GeV/$c$ in the c. m. frame. The numerical fit of Ref. \cite{glasser} is compared to the phase shift $S_{01}$ with the filled (empty) circle, the $P_{01}$ with the filled (empty) square, and the $P_{03}$ with the filled (empty) diamond, respectively. The filled (empty) data point corresponds to the $K^+n\to K^+n\,(K^+n\to K^0p)$ channel.
Given the common negative isovector phase $S_{11}$, set I predicts the dominance of the $P_{03}$ wave over $P_{01}$, while set II says that the dominance between the two is reversed. In set II our fits for the phase shift are consistent with those of Ref. \cite{glasser} except for the $S_{01}$.
} \label{fig3}
\end{figure}

In the case of the isoscalar amplitude, however, the anisotropy appears in the region of ${\rm P_{Lab}}\leq 800$ MeV/$c$, though not shown in the differential cross sections in Fig. \ref{fig3} (left). Thus the $p$-wave as well as the $s$-wave should be included in the phase shift for partial waves $S_{01}$, $P_{01}$, and $P_{03}$.

\subsubsection{The set I}

Including the $p$-wave with the $k^3$ term for the anisotropic angular distribution
the $s$- and $p$-wave phase shifts from the fitting procedure to Damerell's data \cite{damerell} on differential cross sections are given by
\begin{eqnarray}\label{fit-1}
&&\delta^0_{0+}(k)=(a_0 +b_0 k_0+ c_0 k_0^3)\times e^{(k-k_0)/m_0}\,, \nonumber\\
&&\delta^0_{1-}(k)=(a_1+b_1 k_0+c_1 k_0^3)\times e^{(k-k_0)/m_0} \,, \nonumber\\
&&\delta^0_{1+}(k)=(a_3 + b_3 k_0)\times e^{(k-k_0)/m_0}
\end{eqnarray}
with $k_0=220$ and $m_0=100$ MeV/$c$ for $k < 220$ MeV/$c$, and
\begin{eqnarray}\label{fit-2}
&&\delta^0_{0+}(k)=a_0 +b_0 k+c_0 k^3\,, \nonumber\\
&&\delta^0_{1-}(k)=a_1+b_1 k+c_1 k^3 \,, \nonumber\\
&&\delta^0_{1+}(k)=a_3 + b_3 k
\end{eqnarray}
for  $220\leq k\leq 590$ MeV/$c$, and
\begin{eqnarray}\label{fit-3}
&&\delta^0_{0+}(k)=(a_0 +b_0 k_1+c_0 k_1^3)\times e^{-(k-k_1)/m_1}\,, \nonumber\\
&&\delta^0_{1-}(k)=(a_1+b_1 k_1+c_1 k_1^3)\times e^{-(k-k_1)/m_1} \,, \nonumber\\
&&\delta^0_{1+}(k)=(a_3 +b_3 k_1)\times e^{-(k-k_1)/m_1}
\end{eqnarray}
for $k > 590$ MeV/$c$ with $k_1=590$ and $m_1=1500$ MeV/$c$, and the parameters $a_i,\,b_i$ and $c_i$ are collected as the set I in Ref. \cite{ky-kn}.
The exponential function is used to decrease outside of the interval.
The continuity of the amplitude between two different momentum regions further constrains the coefficients $a_i$, $b_i$, and $c_i$ by the boundary condition.

\subsubsection{The set II}

Giacomelli's data \cite{giacomelli73} are used to fix the parameters of set II \cite{ky-kn}.
As before, the phase shifts $\delta^0_{0+}(k)$, $\delta^0_{1-}(k)$ and $\delta^0_{1+}(k)$ are expressed as the same with those in Eq. (\ref{fit-1}) for $k < 335$ MeV/$c$ with $k_0=335$ and $m_0=50$ MeV/$c$, and as in Eq. (\ref{fit-2}) for  $335\leq k \leq 540$ MeV/$c$, and as in Eq. (\ref{fit-3}) for $k > 540$ MeV/$c$ with $k_1=540$ and $m_1=3000$ MeV/$c$, respectively. 

With each parameter set $a_i$, $b_i$, and $c_i$ detailed in Ref. \cite{ky-kn}, Fig. \ref{fig3} (right) displays the phase shift $\delta^{0(1)}_{l+s}$ resulting from the set I fitted to Damerell's data (left) and the other from the set II fitted to Giacomelli's data (right).
Based on the common $S_{11}$ amplitude, we obtain two sets for the isoscalar amplitudes and display $S_{01}$, $P_{01}$ and $P_{03}$ in Fig. \ref{fig3} (right), which nevertheless differ from each other in the same scattering channels.
Our fits from set II are similar to those in Ref. \cite{Sibirtsev}.

Differential and polarization cross sections for elastic and charge exchange $K^+n$ interactions are reproduced in Fig. \ref{fig3} (left), based on the phase shifts from the sets I and II, respectively.
The differential cross sections at ${\rm P_{Lab}}=434$ MeV/$c$ measured by Damerell {\it et\ al.} for both channels are particularly valuable because they provide a testing ground to search for the evidence for the $\Theta^+$ baryon of mass 1535 MeV at the pole position in these reactions.
For the meson-baryon scattering the polarization is given by \cite{giacomelli74}
\begin{eqnarray}\label{pol}
P={2 \textrm{Im}(f g^*)\over |f|^2+|g|^2}
\end{eqnarray}
with the $f$ and $g$ amplitudes as previously discussed.
It is also interesting to note that the polarization of the $K^+n\to K^0p$ reaction is positive, while it is negative in the case of the $K^+n\to K^+n$ reaction \cite{Nakajima}.
These trends continue up to ${\rm P_{Lab}}\approx$1500 MeV/$c$ \cite{ray,Robertson,Watts}. The polarization from set II agrees well with the data, whereas the parameters in set I lead to a result that contradicts the polarization with experiments.
Therefore, the observed polarization serves as a criterion for evaluating the parameters between the two sets.

\subsection{Feasibility of pentaquark $\Theta^+$ baryon}

The BW parameterization for a resonance is applied to a description of the exotic $\Theta^+$ as outlined in Ref. \cite{ky-pin}
\begin{eqnarray}\label{bw-f}
&&f(s,\theta)={1\over k}\sum_Rc_R{(J_R+1/2)\over \epsilon_R-i}e^{-d\epsilon_R^2}P_l(\cos\theta),\\
&&g(s,\theta)={1\over k}\sum_Rc_R{(-1)^{J_R-l+1/2}\over \epsilon_R-i}e^{-d\epsilon_R^2}{dP_l(\cos\theta)\over d\cos\theta}\,,\label{bw-g}
\end{eqnarray}
where $c_R=I_RX_R$ represents the Clebsch-Gordon (CG) coefficient and the elasticity $X_R$. The exponential term with the dimensionless cutoff $d$ is applied for the decrease of a resonance tail.

\begin{figure}[]
\vspace{0.8cm}
\centerline{
\includegraphics[width=6cm]{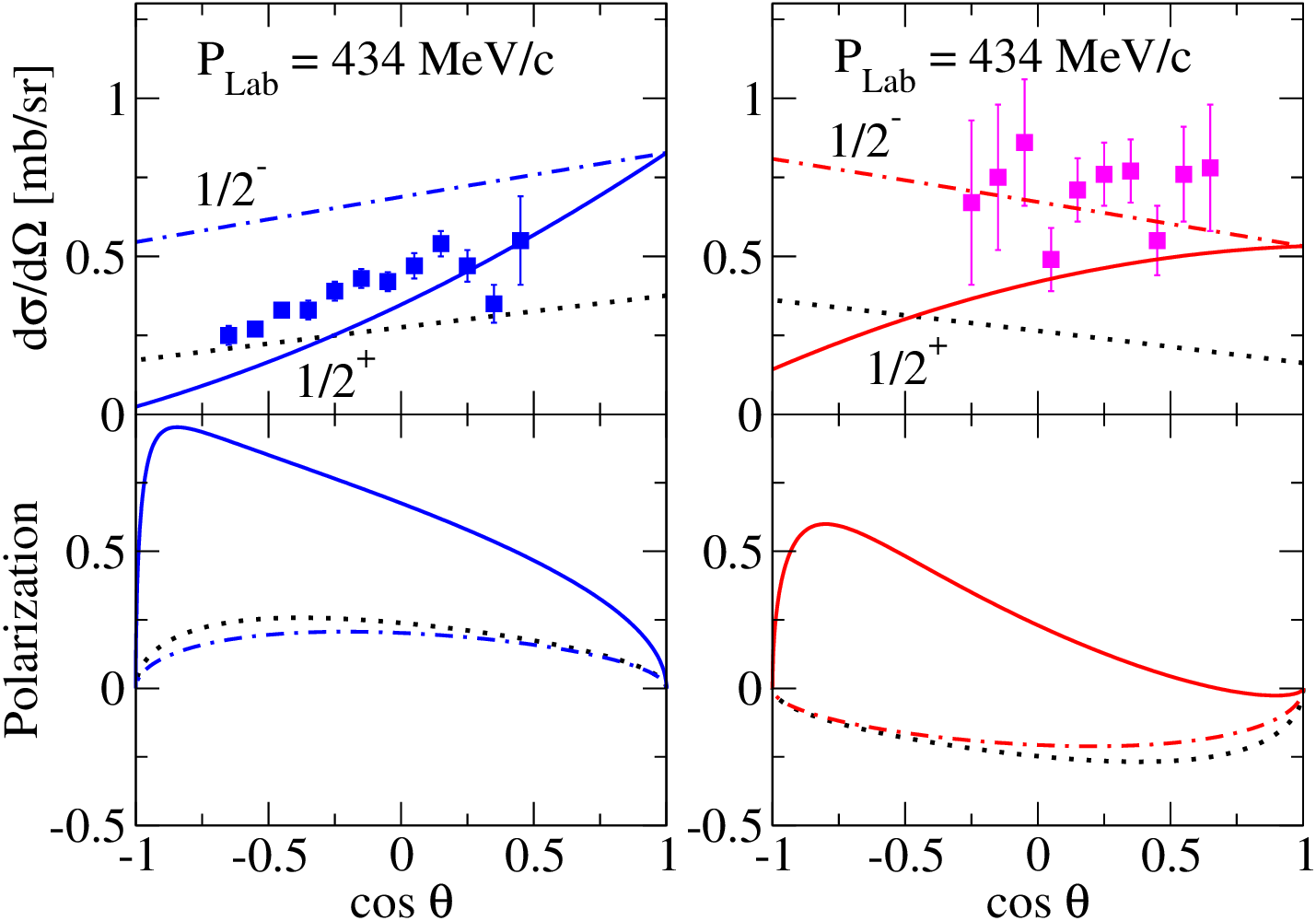}\\
\hspace{0.5cm}
\includegraphics[width=6cm]{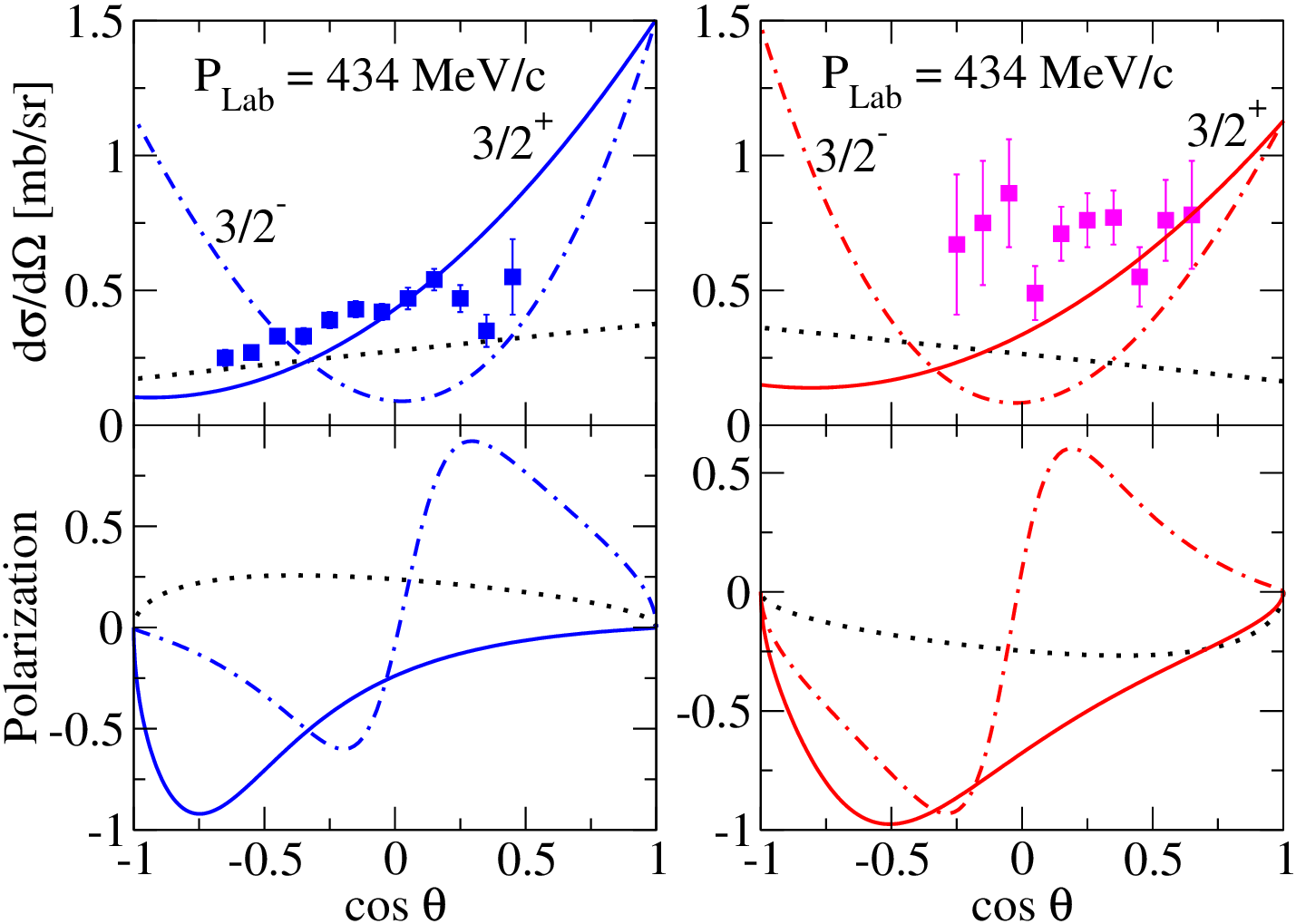}
}
\caption{Role of the $\Theta^+$ in the differential and
polarization observables for $K^+n\to K^0p$ and $K^+n\to K^+n$ at ${\rm P_{Lab}}=$434 MeV/$c$.
The dotted curve corresponds to the phase shift from the set II in Fig. \ref{fig3}, while the solid (dash-dotted) curve represents the full cross section including the $\Theta^+$ baryon of $1/2^+$ and $3/2^+$ ($1/2^-$ and $3/2^-$). Polarization demonstrates a high sensitivity that allows to discern the spin-parity of the $\Theta^+$ among $1/2^\pm$ and $3/2^\pm$.
} \label{fig4}
\end{figure}
\begin{figure}[] 
\vspace{0.8cm}
\centerline{
\includegraphics[width=8cm]{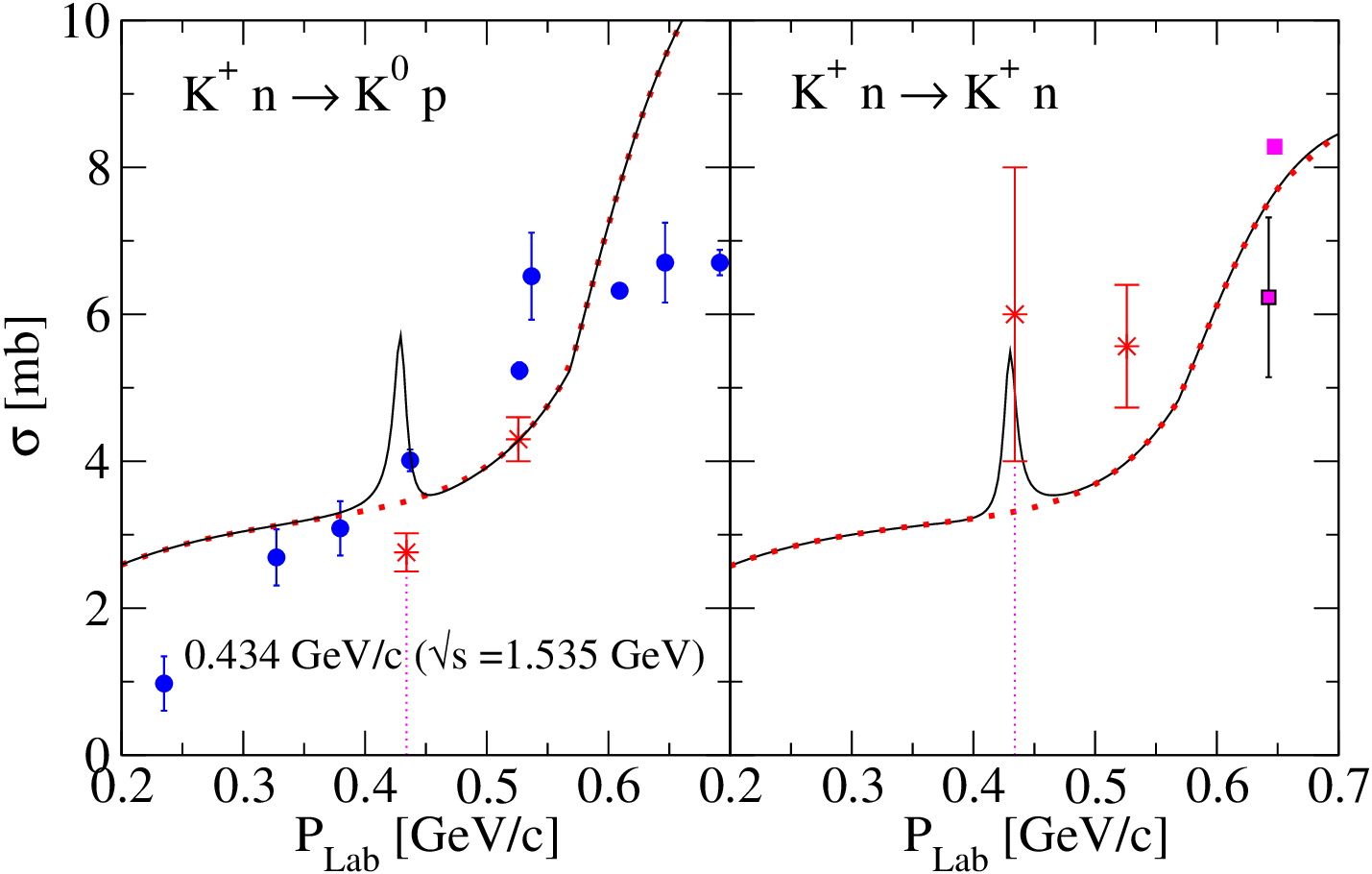}\\
} \caption{Total cross sections for $K^+n\to K^0p$ (left) and
$K^+n\to K^+n$ (right) reactions from set II. The two points at ${\rm
P_{Lab}}=434$ and 526 MeV/$c$ denoted by the cross symbol with
vertical error bars in each cross section are obtained by integrating
the differential cross section data from Ref. \cite{damerell}. In each
cross section, the $\Theta^+$ peak of about 2 mb is shown as an example of
the spin parity $1/2^+$ with parameters as discussed in the text.
Data are taken from Refs. \cite{damerell,giacomelli73}.} \label{fig5}
\end{figure}

Figure \ref{fig4} demonstrates our model prediction for the contribution of the exotic $\Theta^+$ baryon in the BW form to the differential and polarization cross sections for the $K^+n\to K^0\,p$ and $K^+n\to K^+n$ reactions based on Damerell's data.
To agree with the differential cross section at ${\rm P_{Lab}}=434$ MeV/$c$ we choose the mass and width of the $\Theta^+$ baryon as $M_{\Theta^+}=1535$ MeV and $\Gamma_\Theta=5$ MeV for all spin parities, $1/2^\pm$ and $3/2^\pm$. The parameters $c_R=0.25/\sqrt{2}$ including the CG coefficient $1/\sqrt{2}$ and $d=0.01$ are used for the BW form in Eqs. (\ref{bw-f}) and (\ref{bw-g}).

It is interesting to observe that the polarization with the $\Theta^+$ of $1/2^+$ contrasts with the negative parity case which shows an almost vanishing role of the $\Theta^+$. Nevertheless, the case of $\Theta^+$ of $3/2^-$ still produces a discernible polarization as much as $3/2^+$, as shown in the right panels.
Therefore, our prediction tells us that observing polarization is a crucial quantity that should distinguish the presence of the $\Theta^+$ with a contrasting feature between its spin and parities.

The total cross sections for the scattering $K^+n\to K^0p$ (left) and $K^+n\to K^+n$ (right) are shown in Fig. \ref{fig5} with the $\Theta^+$ of $1/2^+$ for illustration purposes, where the solid and dotted curves represent the cross section with and without the $\Theta^+$ in both reactions.

\section{$KN$ scattering on hydrogen target}

Our aim in this section is to introduce the charge exchange and inelastic $K^0\,p$ scatterings, which are free of uncertainties such as the deuteron form factors discussed above.
By providing a theoretical model that describes the existing experimental data for these reactions, we can gain insight to identify the $\Theta^+$ baryon in the reaction on the proton target induced by the neutral kaon beam.

\begin{figure}[b]
\centerline{
\includegraphics[width=6cm]{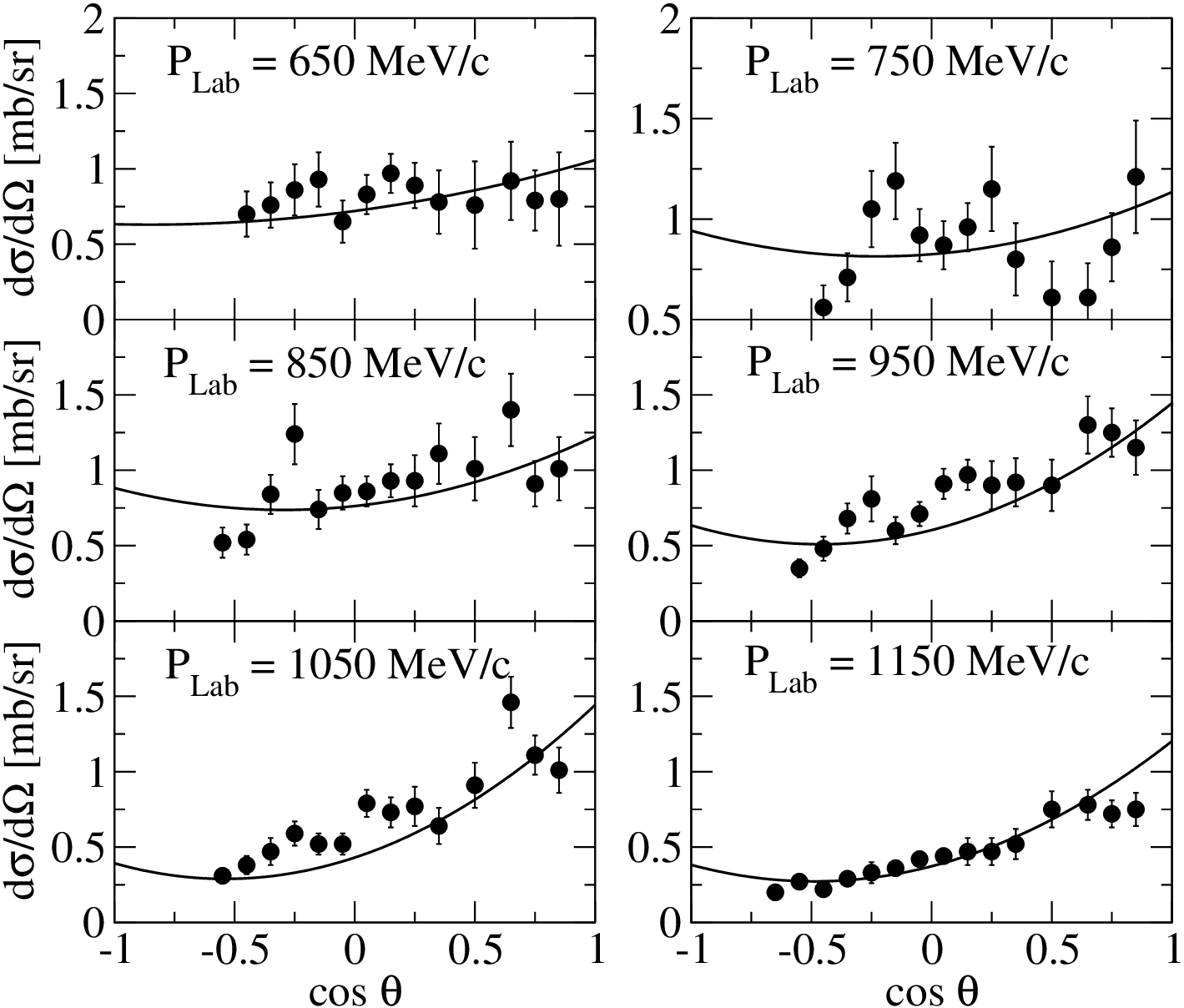}\\
\hspace{0.2cm}
\includegraphics[width=6.2cm]{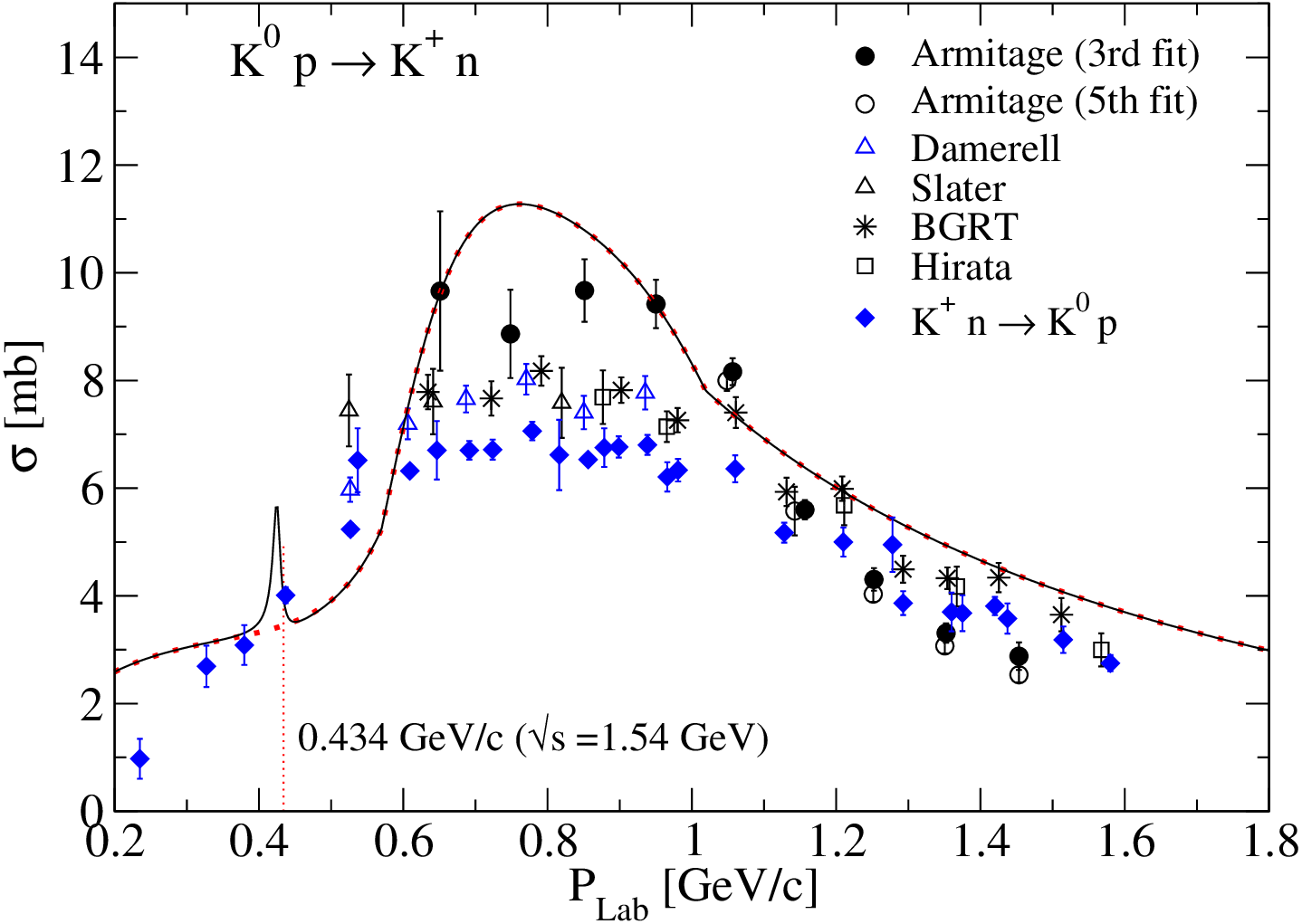}
}
\caption{ Differential and total cross sections for $K^0p\to K^+n$
scattering. Differential cross sections are shown for the kaon momentum
ranging from ${\rm P_{Lab}}=650$ to 1150  MeV/$c$.
The momentum-dependence of the total cross section up to ${\rm P_{Lab}}\approx$ 1.6 GeV/$c$ follows the
exponential terms in Eq. (\ref{fit-3}) working over the region ${\rm P_{Lab}}\approx600$ MeV/$c$.
Total cross section for the $K^+n\to K^0p$ reaction denoted by the filled diamond is presented for comparison.
A similarity is seen in total cross sections of the two channels. Data on differential cross section are taken from Ref. \cite{armitage} and total cross section from Refs. \cite{hirata,damerell,giacomelli72,armitage,slater}.} \label{fig6}
\end{figure}

Let us now investigate the reaction mechanism of the $K^0\,p$ channels in Eqs. (\ref{k0pcex-amp}) and (\ref{k0p-amp}), considering their isoscalar and isovector amplitudes to share with those of the $K^+n$ processes in Eqs. (\ref{k+n-amp}) and (\ref{k+ncex-amp}).

\subsection{$K^0_L \,p \to K^+ \,n$}

Following time reversal between $K^0\,p\to K^+n$ and $K^+n\to K^0\,p$ processes, we employ the phase shift of the set II for the $K^+n\to K^0\,p$ reaction to reproduce the differential and total cross sections for the $K^0\,p\to K^+n$ reaction.

Figure \ref{fig6} displays our model prediction for the differential and total cross sections using the data from Ref. \cite{armitage}. In the total cross section, the peak of the $\Theta^+$ of about 2 mb appears at ${\rm P_{Lab}}=434$ MeV/$c$ with the mass and decay width chosen as before. It is significant to see the validity of time reversal between the two processes by the agreement of the phase shift in Eq. (\ref{k+ncex-amp}) with the experimental data from the third-order fit by Armitage in Eq. (\ref{k0pcex-amp}).
However, the results of the fifth-order fit and other experiments are more consistent with the $K^+n\to K^0\,p$ scattering data, indicating a deviation of the solid curve from the experiment by the fifth-order fit. It is currently difficult to determine which result is closer to reality; if the third-order fit with our prediction is correct, then the validity of time reversal between the fifth-order fit and the $K^+n$ total cross section data that suffers from uncertainties in the deuteron target would be questionable. Otherwise, the analysis of the third-order fit might be inadequate to address the initial $K^0_L$ beam in the weak eigenstate in the reaction. This issue needs to be further investigated in future upgraded experiments.

\subsection{$K^0_L \,p \to K^0_S \,p$}

\begin{figure}[]
\centerline{
\includegraphics[width=6cm]{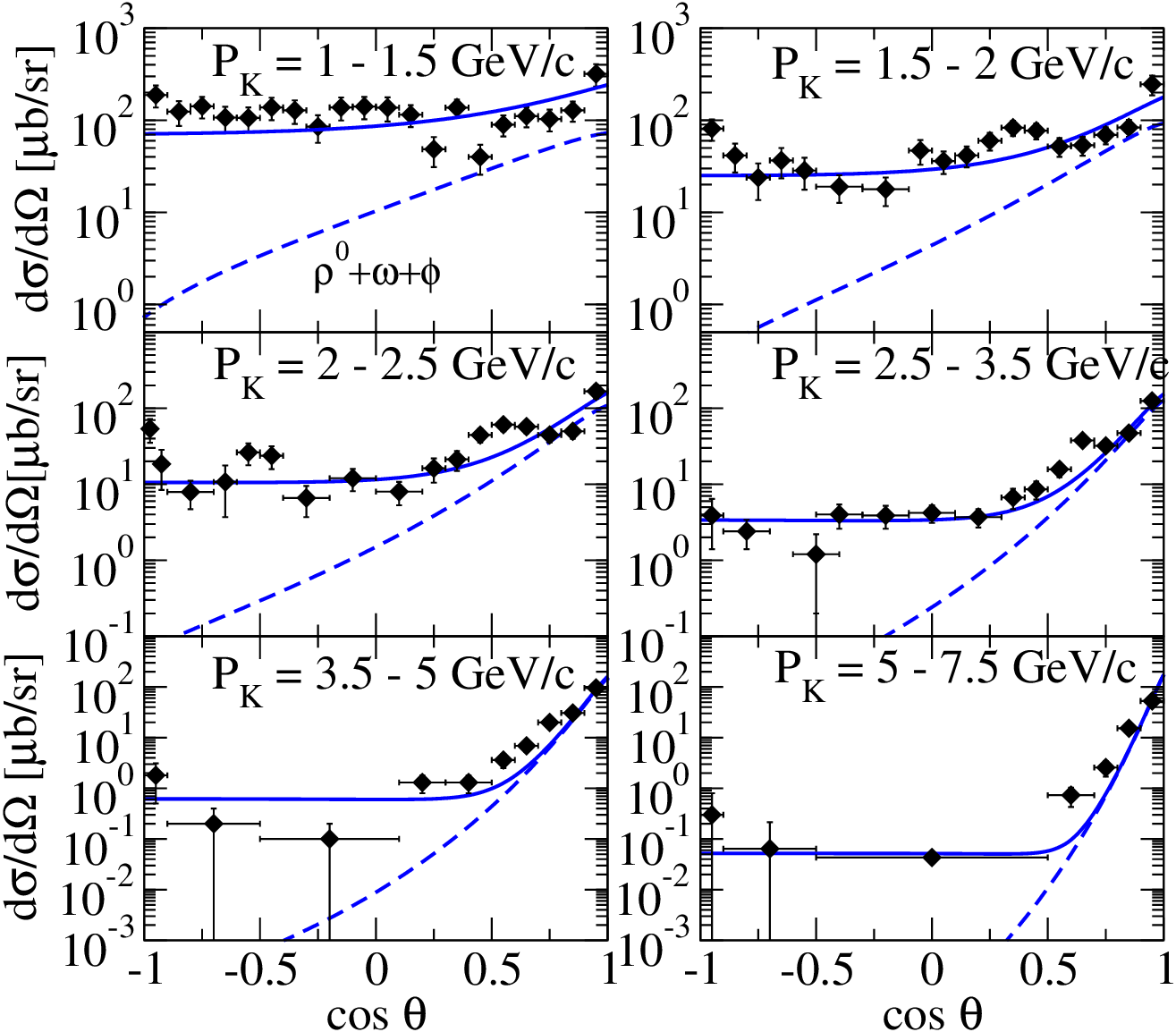}\\
\hspace{0.2cm}
\includegraphics[width=6.2cm]{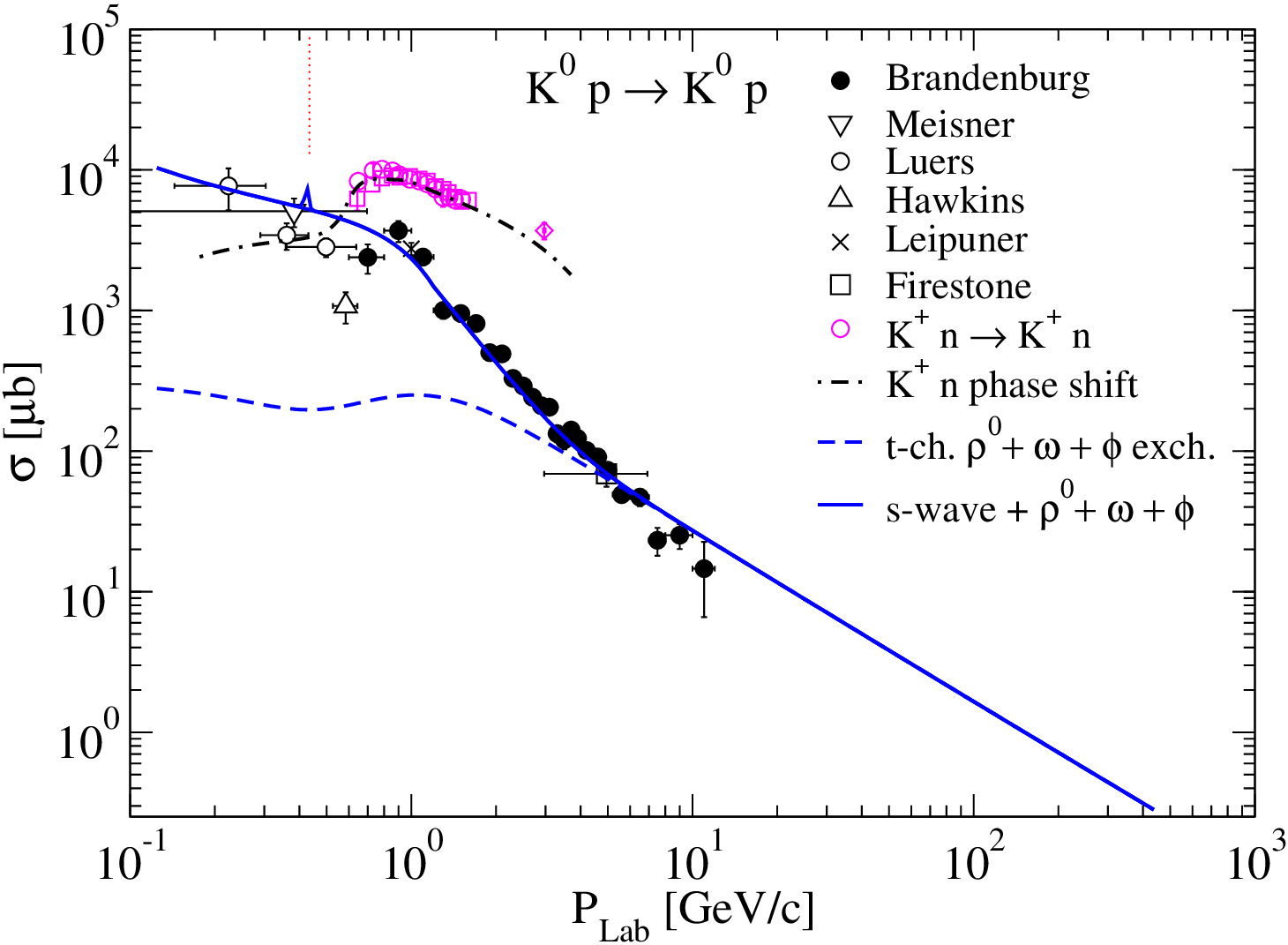}
}
\caption{ Differential and total cross section for the inelastic $K^0\, p\to K^0\,p$
scattering. Total cross section for $K^+n\to K^+n$ channel depicted by the dash-dotted curve is presented for comparison.
The notation for the solid, dashed and dash-dotted curves is given in the text below. The peak of $\Theta^+$ of about 2 mb is located at ${\rm P_{Lab}}$=434 MeV/$c$, as indicated by the vertical line. Data on differential cross section are from Ref. \cite{brandenburg} and total cross section from Refs. \cite{brandenburg,meisner,luers,hawkins,leipuner,firestone}.} \label{fig7}
\end{figure}

Total and differential cross sections for the $K^0_L\,p\to K^0_S\,p$ scattering were measured at SLAC in the momentum range 1$\leq {\rm P_{Lab}}\leq $ 10 GeV/$c$ \cite{brandenburg}.
In Fig. \ref{fig7} (right) it is clear from the shape and size of the total cross section for $K^0_L\,p\to K^0_S\,p$, and $K^+n\to K^+n$ channel given for comparison that the assumed similarity from isospin invariance does not hold between the two reactions.
Therefore, in contrast to the three reactions discussed earlier, we need to consider a new model for the reaction, where the $s$-wave phase shift is revised to fit the low momentum data while taking into account the $t$-channel meson exchange for the description of high momentum behavior.

At high momenta ${\rm P_{Lab}}\geq$ 3 GeV/$c$, analysis of the data favors the Regge pole exchange of light vector meson $\rho^0(775)+\omega(782)$ in the $t$-channel \cite{brandenburg}. Here, we investigate the role of the $\rho^0+\omega+\phi(1020)$ Regge pole exchange, choosing $g_{\rho KK}=2.975$ and $g_{\omega KK}=0.09$ with the Regge trajectories as in the previous $\pi N$ scattering \cite{ky-pin} and the constant phase(=1) for all vector mesons. For the $\phi$ exchange we use $g_{\phi KK}=2.23$ and $g^{v}_{\phi NN}\,(g^{t}_{\phi NN})=-3.8\,(0)$ modified from Ref. \cite{yk-kn} for better agreement.

From threshold to ${\rm P_{Lab}}\approx1.2$ GeV/$c$ we construct the $s$-wave phase shift with the kaon momentum $k$ in the laboratory frame,
\begin{eqnarray}
\delta_0(k)=(7k^2-13k-1)+i(15k^2-26k-2), \ \ \ (\,k\leq1.2\ {\rm GeV}/c\,)
\end{eqnarray}
which is then added to the contribution of the $t$-channel vector meson exchange. The coefficients are fitted to the cross section data as in Eq. (\ref{phase-s11}). For smooth continuity with the meson exchange at $k=1.2$ GeV/$c$ from the high momentum cross section, the exponential factor exp${[-(k-k_0)/m_0]}$ is applied to $\delta_0(k)$ with $k_0=1.2$ GeV/$c$ and $m_0=2$ GeV/$c$.

In Fig. \ref{fig7}, the dashed curve in the differential and total cross sections results from the $\rho^0+\omega+\phi$ exchange describing high momentum region, while the $s$-wave phase shift mounted on the $t$-channel vector meson exchange reproduces the cross section at low momenta.
Thus, our model for the $K^0_L\,p\to K^0_S\,p$ scattering could provide a theoretical tool to study the reaction with the differential and polarization observables at the kaon momentum close to the $\Theta^+$ formation.

\begin{figure}[]
\centerline{
\includegraphics[width=6cm]{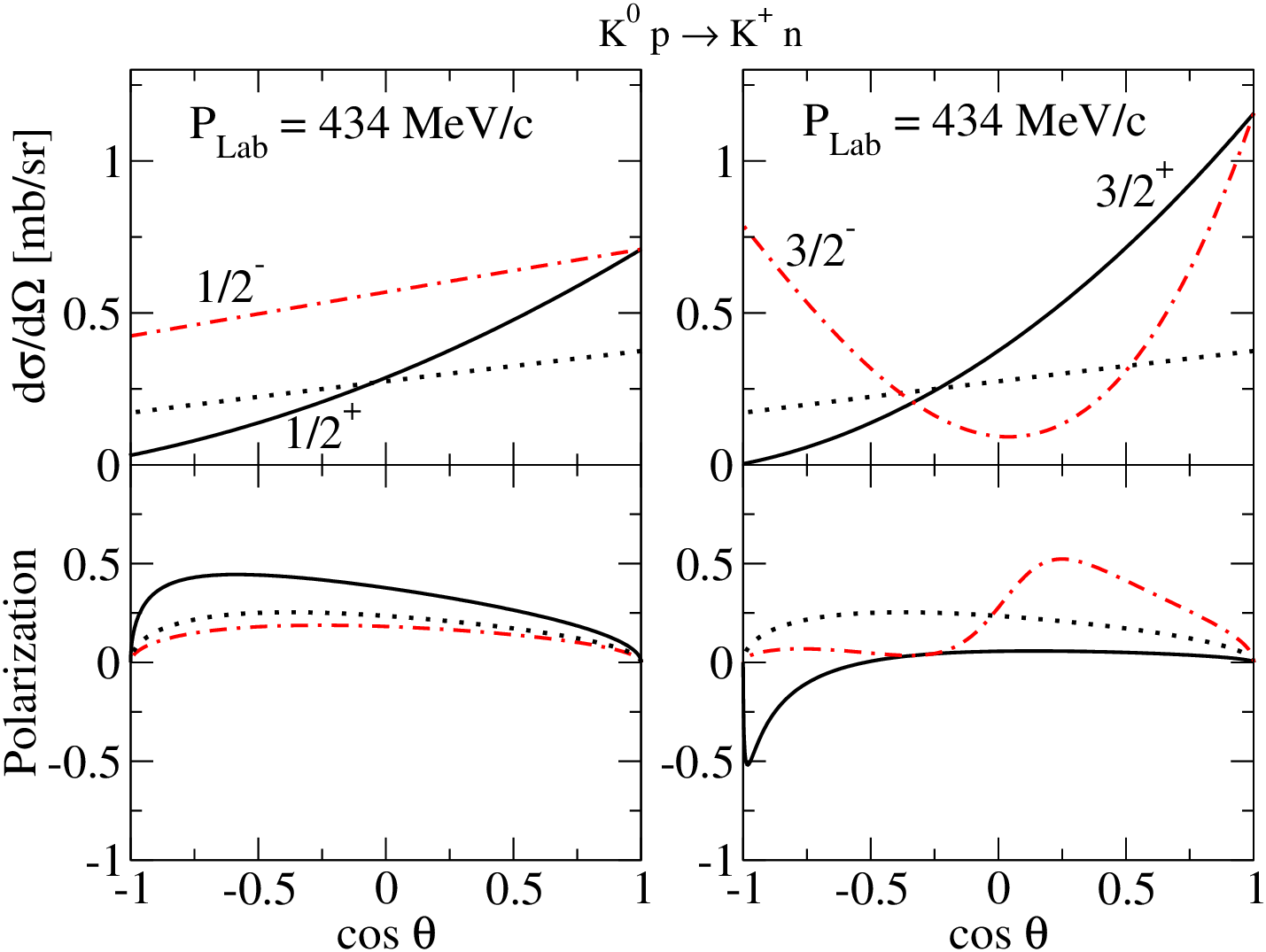}\\
\hspace{0.2cm}
\includegraphics[width=6cm]{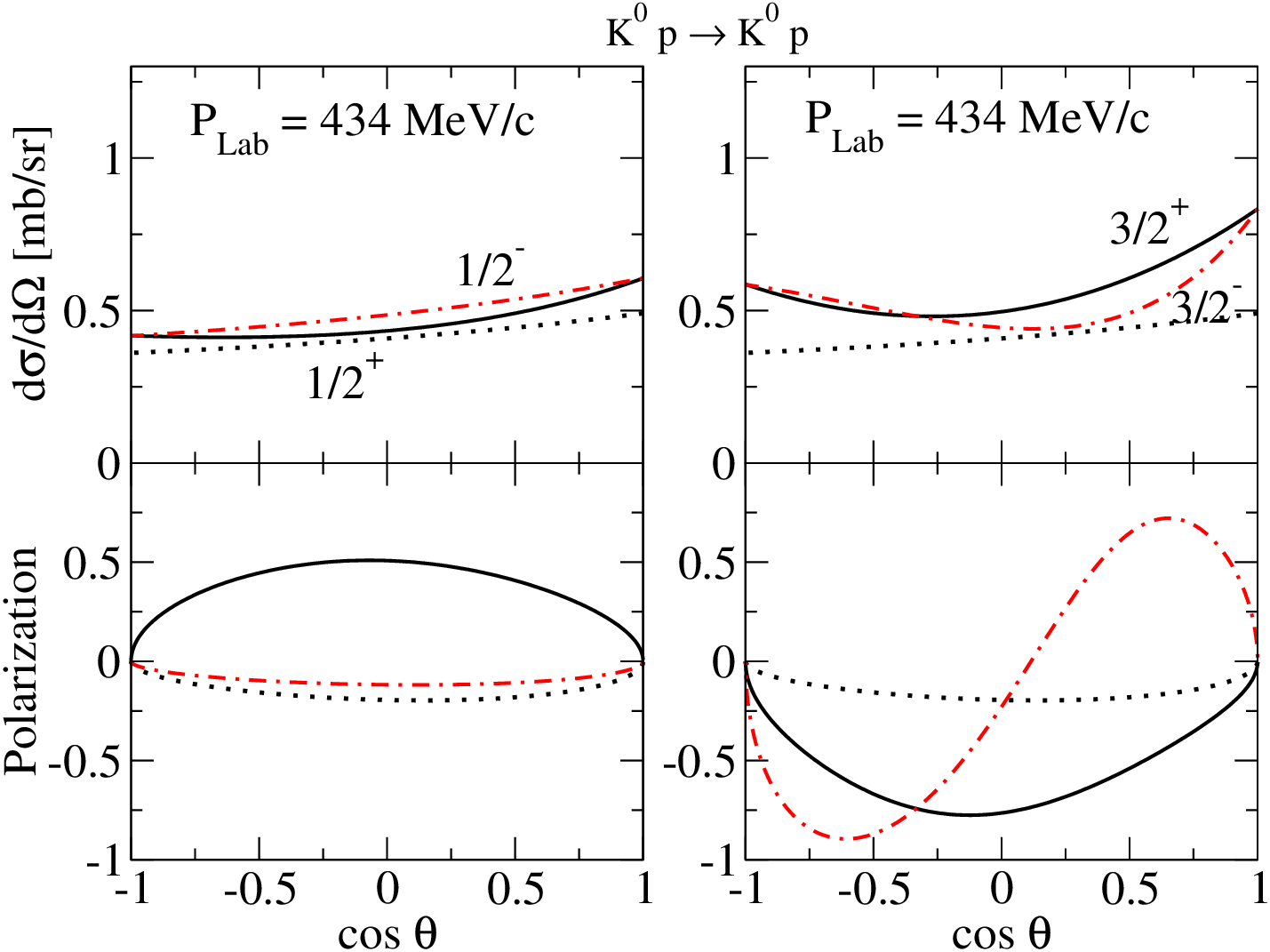}
}
\caption{Role of the $\Theta^+$ in the differential and polarization cross sections for the $K^0\, p\to K^+\,n$ (left) and $K^0\, p\to K^0\,p$ scatterings (right) at ${\rm P_{Lab}}=434$ MeV/$c$.
The dotted curve represents the cross section without $\Theta^+$. The solid and dash-dotted curves correspond to the case of the full cross section with the $\Theta^+$ of $J^P=1/2^+$ and $1/2^-$, and similarly $3/2^+$ and $3/2^-$, respectively.} \label{fig8}
\end{figure}

Note that this reaction is unique compared to the elastic $K^+n$ scattering. It is plausible that the apparent difference between the two reactions arises because the $K^0_L\,p\to K^0_S\,p$ process is not elastic \footnote{There is no $Pomeron$ exchange at high momenta in the inelastic channel.}. Furthermore, the isospin invariance between them holds only when the neutral $K^0$ is in the strong interaction eigenstate. Hence, it is obvious that the $t$-channel meson exchange for the former reaction cannot explain the reaction mechanism of the $K^+n$ elastic process at high momenta.
In our previous study \cite{yk-kn}, we recall that the latter reaction is described by the exchange of $f_0(980)-a_0(980)-\phi(1020)+f_2(1270)-a_2(1320)+Pomeron$ in the forward direction, but without the $\rho^0+\omega$ vector meson exchanges, since they are incapable of $K\bar{K}$ decay for on-mass shells.
Nevertheless, we still find it supporting the current approach to the $K^0_L \,p\to K^0_S \,p$ reaction that the Julich meson exchange model includes $\rho^0$, $\omega$ and $\sigma$ meson couplings with $K\bar{K}$ in $KN$ reactions \cite{buttgen}.
More recently, the Chinese hadron research group has also utilized the modified coupling constants from the Julich model \cite{wu}.

Equipped with these tools, it is now worth searching for evidence of the exotic $\Theta^+$ in the reactions $K^0\,p\to K^+n$ and $K^0\,p\to K^0\,p$ at low momenta.
Differential and polarization cross sections at ${\rm P_{Lab}}=434$ MeV/$c$ are predicted in Fig. \ref{fig8} to distinguish the role of the $\Theta^+$, if possible, in these reactions by varying the possible spin parities.

\section{Summary and conclusion}

We have investigated the possibility of the pentaquark $\Theta^+$ baryon in the four channels of $KN$ elastic, inelastic, and charge exchange scatterings. We begin with the study of the $K^+N$ scattering on the deuteron target, where the uncertainty caused by the deuteron form factors is discussed in the extraction of the experimental data at forward and backward angles.

In studying the reactions $K^+n \to K^0\,p$ and $K^+n \to K^+n$ on the deuteron target, we observe that both share common isovector and isoscalar components in their isospin structure. To address this, we utilize the isovector amplitude from the $K^+p \to K^+p$ scattering to establish that component in these channels.
The remaining isoscalar component is then constructed using the $s$- and $p$-wave phase shifts in the partial wave expansion, fitting it to low momentum data.
Two sets of parameters for the $s$- and $p$-wave phase shifts are tested by fitting the coefficients of the phase shift to experimental data below the region ${\rm P_{Lab}}\leq 800$ MeV/$c$.
For the $K^0\,p\to K^+n$ channel, we apply the $K^+n\to K^0\,p$ scattering amplitude based on the principle of time reversal, achieving good agreement with empirical data.

The exotic $\Theta^+$ is considered using the BW form \cite{ky-pin} to explore its role in the total, differential, and polarization cross sections at the expected momentum ${\rm P_{Lab}}=434$ MeV/$c$, given the $\Theta^+$ mass 1535 MeV. To reproduce Damerell's data for the $K^+n\to K^0\,p$ and $K^+n\to K^+n$ reactions, we choose the decay width of the $\Theta^+$ to be 5 MeV with the elasticity 0.25 and damping factor $d=0.01$ for a rather broader BW tail. We have observed that the peak of the $\Theta^+$ in the total cross section is approximately 2 mb for the $J^p=1/2^+$ case, which is consistent across all four channels. To provide a guide to the spin-parity question of the $\Theta^+$ we test the possible configurations of spin-parity $1/2^\pm$ and $3/2^\pm$. Our findings indicate that polarization is a more effective observable for discriminating the spin-parity, although its size depends on both the elasticity and the decay width of the $\Theta^+$.

In the final section, we introduce the $KN$ scattering on the hydrogen target. We present analyses of the $K^0_L\,p\to K^+n$ and $K^0_L\,p\to K^0_S\,p$ scattering with the prediction for a possible role of the $\Theta^+$ in the differential and polarization cross sections at ${\rm P_{Lab}}=434$ MeV/$c$. It is likely that there is a similarity between the $K^0\,p\to K^+n$ and $K^+n\to K^0\,p$ reactions expected from the validity of time reversal not only in the role of $\Theta^+$ but also in empirical data on both reactions. However, between the two reactions $K^0\,p\to K^0\,p$ and $K^+n\to K^+n$, no similarity is observed either in experimental data on the total cross section or theoretical description of the reaction mechanism, even though isospin invariance is assumed. In fact, the neutral $K^0$ in the reaction is not in the strong interaction eigenstate, which prohibits the naive expectation of the isospin invariance aforementioned.

In conclusion, the current model predictions in Figs. \ref{fig4} and \ref{fig8} for the role of the exotic $\Theta^+$ suggest that the polarization is highly sensitive to the spin parity of the $\Theta^+$ in four channels. This finding makes it easier to distinguish the $\Theta^+$ by observing the polarization rather than the differential and total cross sections in experiments.

\section*{Dedication}

This paper is dedicated to Dmitri Diakonov, Victor Petrov, Maxim Polyakov.
We are indebted to the authors for their invaluable contribution to our profound comprehension of the internal structure of hadrons and quantum chromodynamics.

       \section*{Acknowledgments}
This work was supported by the National Research Foundation of
Korea Grant No. NRF-2022R1A2B5B01002307.



\begin{thebibliography}{10}

\bibitem{gellmann} M. Gell-Mann,
Phys. Lett. {\bf 8}, 214 (1964).

\bibitem{diakonov} D. Diakonov, V. Petrov, and M. V. Polyakov,
Z. Phys. A {\bf359}, 305 (1997).

\bibitem{prasz} M. Prasza{\l}owicz, Phys. Lett. B {\bf575}, 234 (2003).

\bibitem{nakano} T. Nakano et al. [LEPS Collaboration],
Phys. Rev. Lett. {\bf91}, 012002 (2003).


\bibitem{bgyu1} B. G. Yu, T. K. Choi, and C.-R. Ji, Phys. Rev. C {\bf70}, 045205 (2004).

\bibitem{bgyu2} B. G. Yu, T. K. Choi, and C.-R. Ji, J. Phys. G: Nucl. Part. Phys. {\bf32}, 387 (2006).

\bibitem{bgrt} G. Giacomelli et al., Nucl. Phys. B {\bf37}, 577 (1972).

\bibitem{hirata} A. A. Hirata et al., Nucl. Phys. B {\bf 30},  157 (1971).

\bibitem{carroll} A. S. Carroll et al., Phys. Lett. B {\bf 45},  531 (1973).

\bibitem{damerell} C. J. S. Damerell {\it et\ al}.,
Nucl. Phys. B {\bf 94}, 374 (1975).

\bibitem{giacomelli72} G. Giacomelli {\it et\ al}., Nucl. Phys. B {\bf42}, 437 (1972).

\bibitem{giacomelli73} G. Giacomelli {\it et\ al}., Nucl. Phys. B {\bf56}, 346 (1973).

\bibitem{giacomelli74} G. Giacomelli {\it et\ al}., Nucl. Phys. B {\bf71}, 138 (1974).

\bibitem{glasser} R. G. Glasser {\it et\ al}., Phys. Rev. D {\bf 15}, 1200 (1977).

\bibitem{sekihara} T. Sekihara, H.-Ch. Kim, and A. Hosaka, Prog. Theor. Exp. Phys.,  063D03 (2020).

\bibitem{jkahn} J. K. Ahn and S. H. Kim,
J. Korean Phys. Soc. {\bf82}, 579 (2023).

\bibitem{armitage} J. C. M. Armitage {\it et\ al}., Nucl. Phys. B {\bf 123}, 11 (1977).

\bibitem{amaryan24} M. Amaryan et al. [KLF Collaboration],
arXiv:2008.08215 [nucl-ex].

\bibitem{brandenburg} G. W. Brandenburg {\it et\ al}., Phys. Rev. D {\bf 9}, 1939 (1974).

\bibitem{amaryan22} M. Amaryan,
Eur. Phys. J. Plus {\bf137}, 684 (2022).

\bibitem{ky-kn} K.-J. Kong and B.-G. Yu, Few-Body Syst {\bf62}, 73 (2021).

\bibitem{aoki} K. Aoki and D. Jido, Prog. Theor. Exp. Phys. {\bf
2017}, 103D01 (2017).

\bibitem{hashimoto} K. Hashimoto, Phys. Rev. C {\bf29}, 1377  (1984).

\bibitem{yk-kn} B.-G. Yu and K.-J. Kong, Phys. Rev. C {\bf100}, 065206 (2019).

\bibitem{edelstein} R. M. Edelstein {\it et\ al.}, Phys. Rev. D {\bf14}, 702 (1976).

\bibitem{goldhaber} S. Goldhaber, {\it et\ al.},
Phys. Rev. Lett. {\bf9}, 135 (1962).

\bibitem{cameron} W. Cameron {\it et\ al}.,  Nucl. Phys. B {\bf78}, 93 (1974).


\bibitem{buchner} K.Buchner {\it et\ al}., Nucl. Phys. B {\bf44}, 110 (1972).

\bibitem{Sibirtsev} A. Sibirtsev, J. Haidenbauer, S. Krewald, and Ulf-G. Mei\ss ner,
J. Phys. G: Nucl. Part. Phys. {\bf32}, R395 (2006).

\bibitem{Nakajima} K. Nakajima {\it et\ al}., Phys. Lett. B {\bf 112}, 75 (1982).


\bibitem{ray} A. K. Ray {\it et\ al}., Phys. Rev. {\bf 183}, 1183 (1969).

\bibitem{Robertson} A. W. Robertson {\it et\ al}., Phys. Lett. B {\bf 91}, 465 (1980).

\bibitem{Watts} S. J. Watts {\it et\ al}., Phys. Lett. B {\bf 95}, 323 (1980).

\bibitem{ky-pin} K.-J. Kong and B.-G. Yu, Phys. Rev. C {\bf98}, 045207 (2018).

\bibitem{slater} W. E. Slater et al., Phys. Rev. Lett. {\bf7},  378 (1961).

\bibitem{meisner} G. W. Meisner and F. S. Crawford, Phys. Rev. D {\bf3}, 2553 (1971).

\bibitem{luers} D. Luers, I. S. Mittra, W. J. Willis, and S. S. Yamamoto,
in Proceedings of the Aix-en-Prcwence Conference
on Elementary Particles, 1961 (Centre d'Etudes
Nucleaires de Saclay, Gif-sur-Yvette, Seine et Oise,
Saclay, France, 1961), p. 235.

\bibitem{hawkins} C. J. B. Hawkins, Phys. Rev. {\bf156}, 1444 (1967).

\bibitem{leipuner} L. B. Leipuner {\it et\ al}., Phys. Rev. {\bf132}, 2285 (1963).

\bibitem{firestone} A. Firestone {\it et\ al}.,  Phys. Rev. Lett. {\bf16}, 556 (1966).

\bibitem{buttgen} R. B$\ddot{u}$ttgen, K. Holinde and J. Speth, Phys. Lett. B {\bf163}, 305 (1985).

\bibitem{wu} F.-Q. Wu and B.-S. Zou,  Chinese Phys. C {\bf32}, 629 (2008).


\end{thebibliography}
\end{document}